\documentclass[fleqn,twoside]{article}
\usepackage{espcrc2}
\usepackage{graphicx}
\title{
$I=2$ Pion Scattering Length from Two-Pion Wave Function
\thanks{presented by N. Ishizuka}
}
\author{
CP-PACS Collaboration : 
S.~Aoki~\address{
Graduate School of Pure and Applied Sciences,
University of Tsukuba,
Tsukuba, Ibaraki 305-8571, Japan
},
M.~Fukugita~\address{
Institute for Cosmic Ray Research,
University of Tokyo,
Tanashi, Tokyo 188-8502, Japan
},
K-I.~Ishikawa~\address{
Department of Physics,
Hiroshima University,
Higashi-Hiroshima, Hiroshima 739-8526, Japan
},
N.~Ishizuka$^{\rm ~a,}$\address{
Center for Computational Sciences,
University of Tsukuba,
Tsukuba, Ibaraki 305-8577, Japan
},
Y.~Iwasaki$^{\rm ~a}$,
K.~Kanaya$^{\rm ~a}$,
T.~Kaneko\address{
High Energy Accelerator Research Organization (KEK),
Tsukuba, Ibaraki 305-0801, Japan
},
Y.~Kuramashi$^{\rm ~a,d}$,
M.~Okawa$^{\rm ~c}$,
A.~Ukawa$^{\rm ~a,d}$,
T.~Yamazaki$^{\rm ~a}$\thanks{
Present address :
RIKEN BNL Research Center,
Brookhaven National Laboratory,
Upton, NY 11973, USA}
and
T.~Yoshi\'{e}$^{\rm ~a,d}$
}
\begin{document}
%
%
\begin{abstract}
We present a report on a calculation of
scattering length for $I=2$ $S$-wave two-pion system
from two-pion wave function.
Calculations are made
with an RG-improved action for gluons and
improved Wilson action for quarks
at $a^{-1}=1.207(12)\ {\rm GeV}$
on
$16^3 \times 80$,
$20^3 \times 80$ and
$24^3 \times 80$ lattices.
We investigate the validity of
necessary condition for application of L\"uscher's formula
through the wave function.
We find that the condition is satisfied
for lattice volumes $L\ge 3.92\ {\rm fm}$
for the quark mass range $m_\pi^2 = 0.273-0.736\ {\rm GeV}^2$.
We also find that
the scattering length can be extracted with a smaller statistical error
from the wave function
than with a time correlation function used in previous studies.
\end{abstract}
\maketitle
%
%
By now the standard procedure employed for calculating
the scattering length of hadrons from lattice QCD is to
use the finite-size method proposed by L\"uscher,
which relates the finite volume shift of energy eigenvalues $\Delta E$
to the scattering lengths~\cite{Luscher}.
The formula reads
$a_0/(L\pi) = - x - A\cdot x^2 - B\cdot x^3 + O(x^4)$,
where $x = \Delta E \cdot  2 m_\pi L^2 / (4\pi)^2$,
and $A = -8.9136$ and $B = 95.985$ are geometrical constants.
The energy shift $\Delta E$ is extracted from the asymptotic time behavior
of two-pion correlation functions.
In this way a number of calculations has been
made for the $I=2$ $S$-wave pion scattering length~\cite{SGK:a0,GPS:a0,KFMOU:a0,AJ:a0,LZCM:a0,JLQCD:a0,CP-PACS:phsh,Juge:a0,DMML:a0,CP-PACS:phsh_full}.

It should be noted that
the condition $R < L/2$ is assumed
for the two-pion interaction range $R$ and the lattice volume $L^3$
in the derivation of the L\"uscher's formula~\cite{Luscher}.
So far there have been
studies of the lattice volume dependence of the scattering length,
but no direct investigation of the interaction range $R$.
It is very important that
we examine the validity of the necessary condition for the L\"uscher's formula
in our current lattice simulations.
This can be done by investigating the two-pion wave functions,
which is one of the purpose of this article.

Once one has access to the wave functions,
one can try to extract the scattering length from them.
This is the second, and perhaps more interesting,
purpose of this article.
Preliminary results of the present work
was presented at Lattice'03~\cite{IY:ppwf}.

Our idea is based on the derivation of the L\"uscher's formula.
L\"uscher found that the two-pion wave function $\phi(\vec{x})$
on a finite periodic box $L^3$
satisfies an effective Schr\"odinger equation
$( \triangle + k^2 ) \phi( \vec{x} )
  = \int {\rm d}^3 y \ U_{k} (\vec{x},\vec{y}) \phi(\vec{y})
$,
where $\vec{x}$ is the relative coordinate of the two pions and
$k^2$ is related to the two-pion energy eigenvalue
through $E=2 \sqrt{ m_\pi^2 + k^2 }$.
The function $U_k (\vec{x},\vec{y})$ is the Fourier transform of
the modified Bethe-Salpeter kernel introduced in Ref.~\cite{Luscher}.
It is non-local and generally depends on energy.

In the derivation of the formula
it is assumed that $\int {\rm d}^3 y \ U_{k} (\vec{x},\vec{y}) \phi(\vec{y})\not=0$
only for $|\vec{x}| < R < L/2$.
In other words the volume has to be sufficiently large,
so that the boundary condition does not distort the structure of the two-pion interaction.
Out of this region the wave function satisfies the Helmholtz equation
$( \triangle + k^2 ) \phi( \vec{x} ) = 0$.
The general solution of Helmholtz equation on a finite periodic box $L^3$
can be written as
\begin{equation}
  \phi( \vec{x} ) =
      \sum{}_{\vec{p} \in \Gamma }
      \ {\rm e}^{ i \vec{p} \cdot \vec{x} } / ( p^2 - k^2 )
\label{Luscher.div_two}
\end{equation}
up to an overall constant, where
$\Gamma = \{ \ \vec{p} \ | \ \vec{p} = (2\pi)/L\cdot  \vec{n} \ , \ \vec{n}\in {\cal Z}^3 \}$.
Expanding the general solution (\ref{Luscher.div_two})
in terms of spherical Bessel $j_l(x)$ and Neumann $n_l (x)$ functions
for $R < |\vec{x}| < L/2$, we obtain
$\phi( \vec{x} ) =
    \alpha_0 ( k ) \cdot j_0 ( k |\vec{x}| ) + \beta_0 ( k ) \cdot n_0 ( k |\vec{x}| )
  + \cdots
$,
where neglected terms are contributions from states with angular momentum
$l \geq 4$.
The expansion coefficients $\alpha_0 ( k )$ and $\beta_0 ( k )$
yield the scattering phase shift in infinite volume by
$\tan \delta_0 ( k ) = \beta_0 ( k ) / \alpha_0 ( k )$.
In particular for the lowest energy state of the two-pion,
it gives the scattering length by
$\beta_0 ( k ) / \alpha_0 ( k ) = a_0 k + {\rm O}(k^3)$.
The constant $\beta_0 ( k )$ is also
related to $\alpha_0 ( k )$ geometrically
and the relation leads to the L\"uscher's formula.
%
%
\begin{figure}[t]
\vspace*{-0.5cm}
\centering\includegraphics[width=7.0cm]{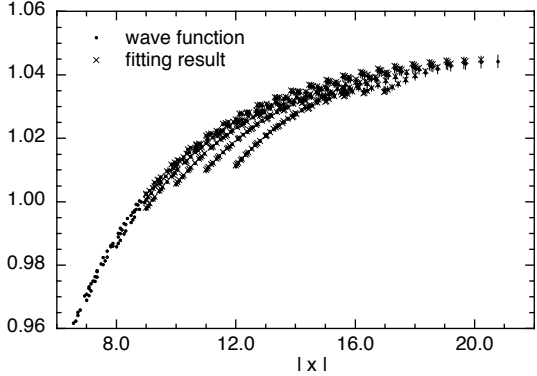}
\vspace*{-1.0cm}
\caption{
\label{FIG.1.fig}
Wave function on $24^3$ lattice at $t=52$ and $m_\pi^2=0.273\ {\rm GeV}^2$.
}
\vspace*{-0.5cm}
\end{figure}
%
%

We define the two-pion wave function by
$\phi( \vec{x}, t ) = \sum_{ {\bf R} , \vec{X}}
     \bigl\langle \pi( R(\vec{x}) + \vec{X},t) \pi(\vec{X},t)
     S  \bigr\rangle$,
where ${\bf R}$ is an element of cubic group, and
summation over ${\bf R}$ and $\vec{X}$
projects out the ${\bf A^{+}}$ sector of the cubic group and that of
the zero center of mass momentum.
In order to enhance signals
we use a source constructed with two wall sources given by
$S=W(t_0) W(t_0+1)$, with $W(t)$ the wall source at $t$,
in Coulomb gauge fixed configurations.

We work in quenched lattice QCD employing an RG-improved action for gluons at
$\beta=2.334$
and an improved Wilson action for quarks at $C_{SW}=1.398$.
The corresponding lattice cutoff is estimated as
$1/a=1.207(12)\ {\rm GeV}$ from $m_\rho$.
The volumes of lattices (number of configuration) are
$16^3\times 80$ $(1200)$,
$20^3\times 80$ $(1000)$ and
$24^3\times 80$ $(506)$ which correspond to
$2.61^3$,
$3.26^3$ and
$3.92^3\ {\rm fm}^3$ in physical units.
Quark masses are chosen to be
$m_\pi^2 = 0.273$, $0.351$, $0.444$, $0.588$ and $0.736\ {\rm GeV}^2$.
Quark propagators are solved with
the Dirichlet boundary condition imposed in the time direction
and the periodic boundary condition in the space directions.
The sources are set at $t_0=12$.

The wave function on a $24^3$ lattice at $t=52$ and $m_\pi^2=0.236\ {\rm GeV^2}$
is plotted in Figure~\ref{FIG.1.fig},
where the horizontal axis is $x=|\vec{x}|$.
We find that the statistical error is very small.
%
%
\begin{figure}[t]
\vspace*{-0.5cm}
\centering\includegraphics[width=7.0cm]{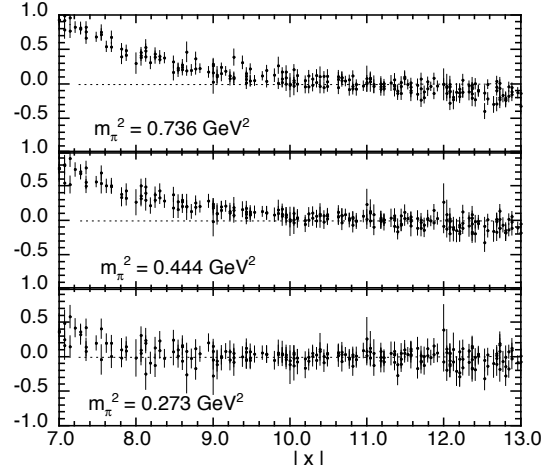}
\vspace*{-1.0cm}
\caption{
\label{FIG.2.fig}
Ratio $U(\vec{x})$ on $24^3$ lattice at $t=52$.
}
\vspace*{-0.8cm}
\end{figure}
%
%

In order to estimate the two-pion interaction range $R$
we construct a ratio
$U(\vec{x}) = ( \triangle + k^2 ) \phi(\vec{x},t) / ( k^2 \phi(\vec{x},t) )$,
where $k^2$ is obtained from the two-pion time correlation function.
This ratio is expected to vanish out of the two-pion interaction range $R < x$.
The ratio $U(\vec{x})$ on a $24^3$ lattice at $t=52$ for several quark masses
are plotted in Fig.~\ref{FIG.2.fig}.
We find a region $U(\vec{x})=0$ for $x < L/2$
for all quark masses within the statistical errors.
The interaction range $R$ tends to be larger for larger quark mass.
We also find that $R \sim 10$ ($1.6\ {\rm fm}$)
in the worst case of our parameters
(the largest quark mass, $m_\pi^2 = 0.736\ {\rm GeV}^2$).
These results mean that the necessary condition for the application
of the L\"uscher's formula is satisfied on the $24^3$ lattice
for all our quark masses.

\begin{figure}[t]
\vspace*{-0.5cm}
\centering\includegraphics[width=7.0cm]{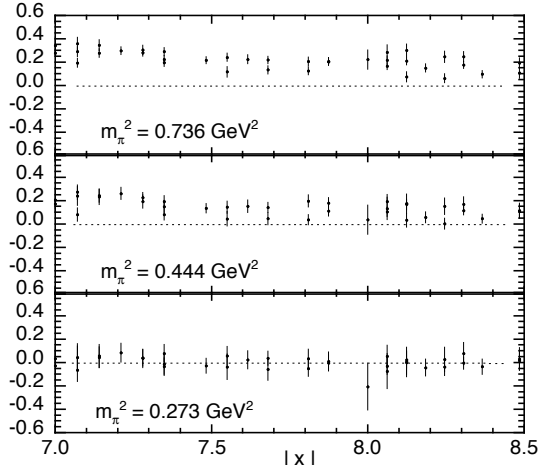}
\vspace*{-1.0cm}
\caption{
\label{FIG.3.fig}
Ratio $U(\vec{x})$ on $16^3$ lattice at $t=52$.
}
\vspace*{-0.8cm}
\end{figure}

The ratio $U(\vec{x})$ on a $16^3$ lattice at $t=52$ for several quark masses
are shown in Fig.~\ref{FIG.3.fig}.
There are deviations of $U(\vec{x})$ from zero for $x < L/2$
for larger quark masses.
These deviation disappear for the lightest quark mass, however.
This trend is also shown for the $20^3$ lattice.
In the following we analyze only data
for which we can clearly find the region $U(\vec{x})=0$ for $x < L/2$.

We estimate the scattering length
by substituting the momentum $k^2$
into the the L\"uscher's formula.
The momentum $k^2$ is obtained by the following two methods.
In one of them
we extract the momentum $k^2$ by a constant fit of the ratio
$V(\vec{x}) = - \triangle \phi(\vec{x},t) / \phi(\vec{x},t)$ for $R < x$.
In the other method
$k^2$ is obtained by
fitting the wave function $\phi(\vec{x},t)$ for $R < x$
with the function (\ref{Luscher.div_two}).
%
%
%
%

An example of fitting the wave function
for a $24^3$ lattice at $t=52$ and $m_\pi^2 = 0.273\ {\rm GeV}^2$
is plotted by cross symbols in Fig.~\ref{FIG.1.fig}.
The fit works very well.

\begin{figure}[t]
\vspace*{-0.5cm}
\centering\includegraphics[width=7.0cm]{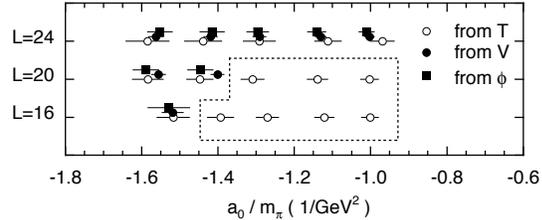}
\vspace*{-1.0cm}
\caption{
\label{FIG.4.fig}
$I=2$ pion scattering length $a_0/m_\pi$ $(1/{\rm GeV}^2)$
obtained from time correlation function (from $T$),
constant fit of $V(\vec{x})$ (from $V$)
and fitting wave function with (\ref{Luscher.div_two}) (from $\phi$).
}
\vspace*{-0.8cm}
\end{figure}

Finally we compare
the scattering length on the three lattice volumes in Fig.\ref{FIG.4.fig}.
We also plot
the results obtained from the two-pion time correlation function
on the $16^3$ and $20^3$ lattices,
for which we can not clearly find the region $U(\vec{x})=0$ for $x < L/2$
(data points enclosed by box in Fig.~\ref{FIG.4.fig}).
We observe that the scattering length obtained from the three methods
are consistent within the statistical errors.
Further the statistical errors of those from our new methods
are smaller than those from the two-pion time correlation function.
We also find no significant volume dependence for all quark masses
including the data points enclosed by the box in Fig.~\ref{FIG.4.fig}.
The necessary condition of the L\"uscher's formula is not satisfied for these data.
However,
the effects of deformations of the two-pion interaction
on the scattering length
due to finite size effect
is apparently small compared with the statistical errors.
%
%

This work is supported in part by Grants-in-Aid of the Ministry of Education
(Nos.
13135204,\hfill\break
13640260, 14046202, 14740173, 15204015,\hfill\break
15540251, 15540279, 15740134, 16028201,\hfill\break
16540228, 16740147 ).
%
%
%
%

%
%
\end{document}